\documentclass[12pt]{article}
\usepackage{graphicx}
\usepackage{amssymb}
\usepackage{epstopdf}
\newcommand{\ignore}[1]{}
\newcommand{\be}{\begin{equation}} \newcommand{\ee}{\end{equation}}
\newcommand{\ba}{\begin{eqnarray}} \newcommand{\ea}{\end{eqnarray}}
 \renewcommand{\bf}{\textbf}
\newcommand{\ra}{\rightarrow}

\newcommand{\p}{\partial}

\def\slasha#1{\setbox0=\hbox{$#1$}#1\hskip-\wd0\hbox 
to\wd0{\hss\sl/\/\hss}}
\def\slashb#1{\setbox0=\hbox{$#1$}#1\hskip-\wd0\dimen0=5pt\advance
        \dimen0 by-\ht0\advance\dimen0 by\dp0\lower0.5\dimen0\hbox
          to\wd0{\hss\sl/\/\hss}}

\input{epsf}

\begin{document}
\title{Supersymmetry \\ and the Lorentz Fine Tuning Problem   }
\author{ Pankaj Jain$^a$ and John P. Ralston$^b$\\
$^a$Physics Department, IIT, Kanpur - 208016, India\\
$^b$Department of Physics \& Astronomy, \\
University of Kansas\\ Lawrence, KS-66045, USA\\
}

\maketitle

\noindent{\bf {Abstract:}}  If physics at the Planck scale requires new 
conceptions of space-time, renormalizable field theories may generically
develop significant violations of Lorentz invariance in the low energy 
sector.  The little recognized ``Lorentz Fine Tuning Problem'' comes 
from logarithmic loop corrections which are not suppressed to the 
enormous extent commonly assumed.  Fine-tuning of parameters at the 
Planck scale is one possible but unpalatable solution.  Here we show 
that violation
of Lorentz invariance is highly suppressed in a supersymmetric theory, 
the Wess-Zumino model. The suppression comes essentially due to the
absence of quadratic divergences in the loop corrections. Since
quadratic divergences are absent in all supersymmetric theories, our 
result is not restricted to the Wess Zumino model.  We also obtain an estimate of the Lorentz violation
expected due to soft supersymmetry breaking. 
We conjecture that supersymmetry may be further 
motivated by yet another fine tuning problem of ordinary quantum field 
theories. 

\vskip 0.8cm
Fine tuning problems are the primary topics in the search for physics 
beyond the Standard Model.  Certain effects of mixing large and small 
scales are naturally suppressed by powers of the respective mass 
ratios, while others are not.  Recently Collins {\it et al} 
\cite{Collins} have highlighted a
new ``Lorentz fine tuning problem''.  The first assumption is that 
space-time ceases to be described by a single background metric at some 
short distance threshold.  Presumably quantum field theory also ceases 
to make sense at short distances, being replaced by a larger framework 
of quantum gravity, implemented by space-time foam, extra-dimensions, 
strings, or something else.  The second assumption is that renormalized 
quantum field theory applies on distances larger than the short 
distance threshold, with the short distance theory serving as a sort of 
boundary condition.  Then we can use field theory to evolve the short 
distance effects up to the laboratory scale. The problem arises for 
renormalization of operators of dimension 4 in 4 dimensions: such 
operators change generically by logarithmic factors, not by power 
suppressions.  With this simple insight, the invariable supposition 
that Lorentz-violating efffects of quantum gravity 
\cite{QG} should be always 
suppressed by enormous ratios \cite{Potting} 
lacks any basis.  Experiments showing no 
large Lorentz violating effects at the low energies of current 
experiments face a contradiction.  One solution is to invoke 
exquisitely precise adjustment of parameters at the quantum gravity 
threshold: the Lorentz Fine Tuning Problem.

We ask whether the virtues of supersymmetry, so acclaimed 
for the gauge hierarchy problem, might usefully confront the threat of 
large Lorentz violation.  In this context ``large'' violation simply 
means an effect of logarithmic order in perturbation theory, which 
would
imply violations at the relative level of $10^{-2}-10^{-4}$, say -- 
very large compared to the tiny values of any phenomenological studies. 
   We find that a broad range of measures of Lorentz violation exhibit 
unexpected cancellations and
have no logarithmic sensitivity in the Wess-Zumino model.  This is 
unexpected and interesting.

The Lorentz Fine Tuning Problem is a much different matter than the 
experimental question of Lorentz symmetry breaking, which has a 
developing phenomenology 
\cite{LorentzViolations,Coleman:1998ti} not discussed here.   
Nor are we concerned with adding new operators of dimension-5 and higher in a 
local theory, which has been discussed in the SUSY 
case \cite{GrootNibbelink:2004za}.  Other work discussing cutoff effects in 
Lorentz violating theories is relevant \cite{Carmona:2000gd}, but we 
are concerned only with supersymmetric theories.  We have not investigated the problem of 
whether all possible models of Lorentz  violation that can be defined 
are suppressed in supersymmetry.  Yet the fact that broad measures 
of Lorentz violation are highly suppressed in one supersymmetric theory 
strongly suggests that supersymmetry may
either solve the Lorentz fine tuning problem, or instruct us towards 
finding theories that are acceptable.

Recall the Wess-Zumino model \cite{Wess}, with Lagrangian density
\be
{\cal L} = {\cal L}_0 + {\cal L}_m + {\cal L}_I
\ee
where
\be
{\cal L}_0 = {1\over 2} \partial^\mu A \partial_\mu A
+ {1\over 2} \partial^\mu B \partial_\mu B
+ {i\over 2} \bar \psi\slasha\partial\psi + {1\over 2}{\cal F}^2 +
{1\over 2}{\cal G}^2
\ee
\be
{\cal L}_m = m[{\cal F}A + {\cal G}B - {1\over 2}\bar \psi\psi]
\ee
\be
{\cal L}_I = {g\over \sqrt 2}[{\cal F}A^2 - {\cal G}B^2
+ 2 {\cal G} AB
-\bar \psi (A - i\gamma^5 B)\psi]
\ee

This model needs regularization and we turn to implementing a 
Lorentz-violating cutoff.  Even if Lorentz symmetry is broken, nothing 
stops one from expanding all fields, momenta, and breaking parameters 
in a complete set of Lorentz representations \cite{Coleman:1998ti}.  
Completeness of representations will organize terms into convenient 
categories whether or not the symmetry is good.  Three basic forms of 
cutoff are a scalar $\Lambda$,  4-vector parameters $\Lambda^{\mu}$ and 
symmetric tensor parameters $\Lambda^{\mu \nu}$.  In a quantity with 
one external momentum $k^{\mu}$,  Lorentz-invariant terms are functions 
of $k^{2}$ and the scalar cutoff, while Lorentz-breaking functions will 
depend on $( k \cdot \Lambda)^{2} $ or $ k\cdot \Lambda \cdot k$.  We 
do not consider functions odd in $k \cdot \Lambda$ which depend on the 
direction of $k^{\mu}$. 
The SUSY invariance of our model would be broken by such odd dependence.
 A cutoff is basically a function 
$f(\p_\mu)$ where $\p_{\mu} = (\p_{0}, \,
\p_{i}) =\p /\p x^{\mu}$ is the derivative operator. We assume
that $f(\p_\mu)$ is a function only of the four variables $(\p_0^2/
\Lambda_0^2,\p_i^2/\Lambda_i^2)$ where
$\Lambda_{\mu}$ are the cutoffs, either 4-vectors or tensor 
eigenvalues.  This model of Lorentz violation is a bit more general 
than the study of Ref. \cite{Collins}.  We need not assume rotational 
invariance, of course.  Regularization with terms involving $k_{0}$ can 
generate unphysical poles, violating causality and standard assumptions 
such as Wick rotation. At the same time not every form of 
regularization is diseased, so we have left the choice of 
regularization open.  

We can now directly calculate the leading loop corrections to the
scalar propagator.  We wish to
retain SUSY invariance in calculations, so we modify the free 
Lagrangian density
${\cal  L}_0 + {\cal L}_m$ by inserting the cutoff function $f$ in 
coordinate space between pairs of
fields. An alternative nonlocal regularization of supersymmetric theories
has also been proposed in 
Ref. \cite {Kleppe}. However in the present paper we are interested in 
only modifying the propagators with a cutoff function, in direct analogy
with the procedure used in Ref. \cite{Collins}. 
Quantization of such formally non-local field theories goes smoothly
in the canonical way \cite{canonical}.  The regulated terms become
\ba
{\cal L}'_0 &=& {1\over 2} \partial^\mu A
f(\partial_\mu) \partial_\mu A
+ {1\over 2} \partial^\mu B
f(\partial_\mu)\partial_\mu B
+ {i\over 2} \bar \psi
f(\partial_\mu)
\slasha\partial\psi\nonumber\\   &+&
{1\over 2}{\cal F}f(\partial_\mu)
{\cal F}+
{1\over 2}{\cal G}f(\partial_\mu){\cal G}\\
{\cal L}'_m &=& m[{\cal F}f(\partial_\mu)A +
{\cal G}f(\partial_\mu)B -
{1\over 2}\bar \psi f(\partial_\mu)\psi]
\ea
It can be verified that the action remains invariant under SUSY 
transformations. 

We may now eliminate the auxiliary fields $\cal F$ and $\cal G$. The Lagrangian
terms needed for our calculation can be written as: \ba
{\cal L}' &=& {1\over 2} \partial^\mu A f(\partial_\mu)
\partial_\mu A
+ {1\over 2} \partial^\mu B f(\partial_\mu)\partial_\mu B
+ {i\over 2} \bar \psi f(\partial_\mu)
\slasha\partial\psi\nonumber\\  &-&
{m^2\over 2}[Af(\partial_\mu)A +
Bf(\partial_\mu)B] -
{m\over 2}\bar \psi f(\partial_\mu)\psi\nonumber\\
&-& {mg\over \sqrt 2} A(A^2 + B^2) - {g\over \sqrt 2}\bar\psi
(A - i\gamma^5 B)\psi\nonumber\\
&-& {g^2\over 4}\Big[A^2f^{-1}(\partial_\mu) A^2
+ B^2f^{-1}(\partial_\mu)B^2
- A^2f^{-1}(\partial_\mu)B^2 \nonumber\\ &-&
B^2f^{-1}(\partial_\mu) A^2 + 4 
ABf^{-1}(\partial_\mu)AB
\Big]
\ea
Note we have not inserted any cutoff function in the
interaction lagrangian.   The
$f(\p_{\mu})$ dependence in the interaction
terms arises entirely due to the elimination of the auxiliary fields.

\begin{figure}
\epsfbox{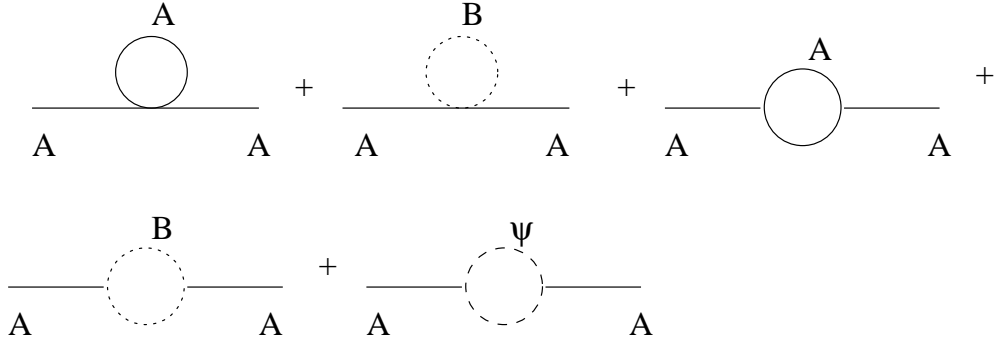}
\caption{The self energy contributions to the scalar particle $A$
at one loop order.  Letter symbols mark the fields. }
\label{fig:diagrams}
\end{figure}

The diagrams that contribute to the self energy $\Pi(k)$ of the scalar
particle with momentum $k$ are given in Fig. \ref{fig:diagrams}.
The contributions add to give:
\be
\Pi(k) = 4 g^2\int {d^4q\over (2\pi)^4} {1\over f(k_\mu+q_\mu)
f(q_\mu)}{7m^2 + k^2 + k\cdot q\over
(q^2-m^2+i\epsilon)[(k+q)^2-m^2+i\epsilon]}\ ,
\label{Eq:Pi}
\ee

There are several possible measures of Lorentz violation. In the low 
energy limit, meaning momenta small compared to $\Lambda_{\mu}$, 
non-trivial dependence starts at second order in $k^{\mu}$.    Collins 
{\it et al} \cite{Collins} consider a quantity $\xi$  defined by a 
combination of second-derivatives:
\be
\xi = \lim_{k \ra 0} \,\left( {\partial^2 \Pi(k)\over \partial(k^0)^2} +
{\partial^2 \Pi(k)\over \partial(k^1)^2}\right)
\ee
We adopt $\xi$ as a measure, since our cutoffs have been arranged to 
maintain the feature that $\xi=0$ when $\Pi =\Pi(k^{2})$, namely if 
Lorentz symmetry exists, with non-zero $\xi$ probing symmetry-breaking. 
  A typical derivative is
\ba
{\partial^2 \Pi(k)\over \partial(k^0)^2}\Bigg|_{k_\mu=0}  & =&
4g^2\int {d^4q\over (2\pi)^4}\Bigg\lbrace{1\over f_q}
\left({\partial^2\over \partial (k^0)^2}
{1\over f_{k+q}}\right){7m^2\over D_q^2}\nonumber\\
&+& {2\over f_q} \left({\partial\over \partial k^0}{1\over 
f_{k+q}}\right)
\left[{q_0\over D_q^2}-{14 m^2 q_0\over D_q^3}\right]\nonumber\\
&+&{1\over f_q^2}\left[
{2\over D_q^2} - {4q_0^2\over D_q^3} + {56m^2q_0^2\over D_q^4}
-{14m^2\over D_q^3}\right]\Bigg\rbrace ,
\label{Eq:partialq0}
\ea
evaluated at $k_\mu= 0$. Here
$D_q = q^2 - m^2 + i\epsilon$, and $f_q = f(q_\mu)$.  Some 
algebra shows that the first term on the right hand side of Eq. 
\ref{Eq:partialq0} is proportional to $(m/ \Lambda)^2$, where $\Lambda$ 
come from the set of $\Lambda_0, \Lambda_i$, times integrals that are 
at most logarithmically divergent.  This term is negligible as $\Lambda 
\ra \infty$.

We simplify the second term in Eq. \ref{Eq:partialq0} by writing
$${2\over f_q}{\partial\over \partial k^0}{1\over f_{k+q}}\Bigg|_{k=0}
= {2\over f_q} {\partial\over \partial q^0}{1\over f_{q}}
= {\partial\over \partial q^0}{1\over f_{q}^2}\ .$$
Integration by parts yields
\be
{\partial^2 \Pi(k)\over \partial(k^0)^2}\Bigg|_{k=0}   =
4g^2\int {d^4q\over (2\pi)^4}{1\over f_q^2} \Bigg\lbrace
{1\over D_q^2}  - 28 m^2  {q_0^2\over D_q^4}\Bigg\rbrace,
\label{Eq:partialq01}
\ee
where the surface term at fixed $\Lambda_{\mu}$ is zero and can be 
dropped.
Similar manipulations lead to the expression,
\be
{\partial^2 \Pi(k)\over \partial(k^1)^2}\Bigg|_{k=0}   =
4g^2\int {d^4q\over (2\pi)^4}{1\over f_q^2} \Bigg\lbrace
-{1\over D_q^2}  - 28 m^2  {q_1^2\over D_q^4}\Bigg\rbrace
\label{Eq:partialq1}
\ee
The measure of Lorenz violation yields a surprising cancellation:
\be
\xi =  4g^2\int {d^4q\over (2\pi)^4}{1\over f_q^2}\left[
-28m^2{q_0^2+q_1^2\over D_q^4}\right] + O(m^2/\Lambda^2)
\ee
The integral on the right hand side is convergent even in the limit
$f_q\ra 1$. Hence we can estimate it by setting $f(q_\mu)=1$ and then
making a Wick rotation. The two terms in the integral cancel by Euclidean
rotational invariance and hence we find that $\xi$ is of order $(m/\Lambda)^2$.
In fact the three terms contributing to $\Pi(k)$,  Eq.
\ref{Eq:Pi}, individually give contributions to $\xi$
which are power suppressed. 
It is rather astonishing to find a {\it finite} term, however small, at low 
energy generated by Lorentz violation as a remnant of quantum gravity.  
It is a new form of observable reminiscent of the anomalous magnetic 
moment of the electron in QED, which by {\it not} appearing at zeroeth 
order, had to be finite in the right theory.

Let us examine the main source of the cancellation. In a generic theory 
without supersymmetry, the scalar self-energy is quadratically 
divergent.
In that event we would get a term of the kind
\be
\Delta\Pi(k) \sim  -4g^2\int {d^4q\over (2\pi)^4} {q^2\over
(q^2-m^2+i\epsilon)[(k+q)^2-m^2+i\epsilon]}\ ,
\label{Eq:DeltaPi}
\ee
The corresponding contribution to $\xi$ would be of the form,
\be
\Delta\xi\sim -4\int {d^4q\over (2\pi)^4} {q^2 (q_0^2 + q_1^2)\over 
D_q^4}\ ,
\ee
which is logarithmically divergent.  This is the source of the $LFT$ 
problem. Such a term does not arise in the parameter $\xi$ in the 
supersymmetric theory due to the absence of quadratic
divergences.  
The mechanism suggests that our result may be more general. 
Indeed we have only assumed that $f(k_\mu)$ regulates the
integral in establishing that $\xi$ is power suppressed and not made any
further assumptions regarding its form. 
Furthermore since quadratic divergences are absent in all supersymmetric 
theories, such as the MSSM, we expect that LFT problem may
be absent in these theories. 

We may also consider other possible measures of Lorentz 
violation which may be formed by taking higher derivatives of $\Pi(k)$ in the 
limit $k\ra 0$. We find that all terms which involve more than
two derivatives of $\Pi(k)$ are convergent even in the absence of 
the cutoff function. Hence all measures of Lorentz violation 
constructed from more than two derivatives will be power suppressed.  
We are, therefore, unable to construct any measure of Lorentz violation which
is not power suppressed. 

We then find that reasonable measures of Lorentz symmetry breaking due 
to loop corrections are highly suppressed due to supersymmetry.   Given 
the mechanism of suppression, it is perfectly justified to conjecture 
that supersymmetry may solve yet another fine tuning problem.  However 
we cannot possibly examine all models in all possible ways.  If 
something goes wrong in some different calculation, there is a 
tantalizing possibility that a special theory might have special 
features bringing some notion of uniqueness in the search for new 
physics.

We conclude with several observations:

$\bullet$ Any physical appeal to SUSY is an appeal to broken SUSY.  If 
SUSY is broken
at a mass scale $M<< \Lambda$, our type of Lorentz-violation would be 
power-suppressed over the broad range, just as commonly imagined before 
the $LFT$ crisis emerged.  One might estimate Lorentz-violating effects 
using the new parameter ratio $(M/\Lambda)^2$, which brings new scales 
into the problem. 
Here we are interested in soft SUSY breaking terms 
which would break the degeneracy between the SUSY partners. In order to
make an estimate we assume that the scalars in our model acquire a 
large mass of the order of SUSY breaking scale $M>>m$. We then consider
the quadratically divergent terms which contribute to $\Pi$, 
\ba
\Pi(k) &\sim & -4 g^2\int {d^4q\over (2\pi)^4}{1\over f_{k+q}f_{q}} {q^2\over
(q^2-m^2+i\epsilon)[(k+q)^2-m^2+i\epsilon]}\nonumber \\
&+& 4g^2\int {d^4q\over (2\pi)^4}
{1\over f_{k+q}f_{q}}{1\over q^2 -M^2 + i\epsilon }\ .
\label{Eq:SUSY_Breaking}
\ea
We may now estimate the contribution to the Lorentz violating parameter
$\xi$ from these terms. The leading contribution, coming from quadratically
divergent terms, will cancel among the two terms on the right hand side
of this equation. The subleading term may be estimated by assuming a
simple model, \be f_q={1\over 1+\vec q^2/\Lambda^2}\ .\ee
Using this model we find that the dominant contribution to $\xi$ is
proportional to $(M/\Lambda)^2\log (M/\Lambda)$. Since the Lorentz
invariance has been tested to order $10^{-20}$, our estimate rules out 
SUSY breaking scales larger than $10^{9}$ GeV, assuming
that $\Lambda$ is of the order of the Planck mass. 
Given this estimate of Lorentz violation, we may
reconsider quantities capable of coherently 
accumulating very minute symmetry-breaking effects \cite{Ralston:2003pf}. 

$\bullet$ It is sensible to ask whether the Lorentz Fine Tuning problem 
might be a disease of perturbative physics, which is poorly suited to 
handling widely separated energy scales.  There is no fine tuning 
problem in condensed matter or molecular physics because methods more 
sophisticated than perturbation theory are used.  Despite popular 
appeal it is not really necessary to invoke SUSY just because Nature 
has big energy scales.  The physical question is whether actual 
physical systems exhibit a vicious interplay of short distance symmetry 
breaking and long distance behavior.  The elastic (acoustic) modes of a 
crystal are a prototype:  while being massless due to a Goldstone 
symmetry \cite{Anderson}, and in the infrared region seemingly 
insensitive to the ultraviolet cutoff, their dispersion relations may 
be highly {\it anisotropic} due to the anisotropy of the smallest unit 
cell.  One cannot escape the Lorentz Fine Tuning problem by calling it 
a pathology of calculational procedures.  Of course condensed matter 
interactions are electromagnetic, leaving open what must happen under 
interactions with different renormalization properties.

$\bullet$ At this time the role of general covariance in microscopic 
physics is entirely unknown.  With nothing but faith it is common to 
extrapolate the macroscopic phenomenology of General Relativity ($GR$) 
in a straightforward way, and assume its symmetries apply down to the 
shortest distance scales.  Suppose there existed a way to calculate the 
evolution of the ``renormalized metric'', whatever that means, {\it 
consistent with the symmetries of general covariance}, all the way down 
to the Planck scale.  Then the choice of the metric in a particular 
calculation reduces to a meaningless choice of gauge. Moreover, if the 
ambitions of $GR$ are met in the most simple way, then one and only one 
metric serves everywhere in one universal way, so that one consistent 
calculation cannot falsify another.  For certain ``minimal-coupling'' 
theories it will not be possible or meaningful {\it locally} to resolve 
whether or not Lorentz violation in a renormalized metric exists.  Yet 
experiments in microscopic physics are very local, integrating over 
tiny distances set by external momenta.  Then the question of Lorentz 
violation would hinge on correlations and fluctuations comparing what 
is meant by the metric in different circumstances.  The notion of a 
simple set of background parameters violating Lorentz invariance is 
utterly inadequate for this task.  Lacking a framework in which to 
carry out such calculations, further prognostication seems unjustified: 
there may be millions of variations, one unique resolution, or no 
resolution possible.   Moreover, the prospect of non-minimally coupled 
Lorentz-violating theories opens a Pandora's Box of new possibilities.

We come to the surprising realization that the scope of the Lorentz 
fine tuning problem may be very broad and deep.  Would experimental 
detection of Lorentz violation rule out general covariance in 
microscopic physics, or simply reject its most naive interpretation?   
Will that class of theories ignoring either SUSY or general covariance, 
meaning all ordinary quantum field theories, confront the Lorentz Fine 
Tuning Problem without defense, and then be ruled out indirectly?  Can 
resolution of $LFT$ point the way towards constructing 
phenomenologically realistic theories?

\bigskip
\noindent
\bf{Acknowledgements:} JP thanks John Collins, Tom DeGrande, and Danny 
Marfatia for discussions, and the organizers of Miami '04 for bringing 
people together.   PJ thanks Satish Joglekar and Sreerup Raychaudhuri 
for useful discussions.  Supported in part under DOE grant number 
DE-FG02-04ER14308.


\begin{thebibliography}{unsrt}
\bibitem{Collins} J. Collins, A. Perez, D. Sudarsky, L. Urrutia
and H. Vucetich, Phys. Rev. Lett. \bf {93}, 191301 (2004); 
gr-qc/0403053; see also e-proceedings of {\it Miami 2004}, in 
preparation.

\bibitem{QG} V. A. Kosteleck\'{y} and S. Samuel, Phys. Rev. \bf{D 39},
683 (1989); G. Amelino-Camelia {\it et al.}, Int. J. Mod. Phys. \bf{A 12},
607 (1997); J. R. Ellis {\it et al.},  Phys. Rev. \bf{D 61}, 027503 (2000);
J. Alfaro {\it et al.}, Phys. Rev. \bf{D 66}, 124006 (2002); H. Sahlmann and
T. thiemann, gr-qc/0207031; R. Gambini and J. Pullin, Phys. Rev. \bf{D 59},
124021 (1999). 

\bibitem{Potting} V. A. Kosteleck\'{y} and R. Potting, Phys. Rev. \bf{D 51},
3923 (1995).

\bibitem{LorentzViolations}  
D.~Colladay and V.~A.~Kosteleck\'{y},
Phys.\ Rev.\ D \bf {58}, 116002 (1998), hep-ph/9809521; 
G. Amelino-Camelia {\it et al.}, Nature (London)
\bf {393}, 763 (1998); 
J. R. Ellis {\it et al.}, Astrophys. J. \bf {535}, 139 (2000);
R. J. Gleiser and C. N. Kozameh, Phys. Rev. D \bf {64}, 083007 (2001); 
C. E. Carlson {\it et al.},  Phys. Lett. \bf {B 518},
 201 (2001);
D. Sudarsky {\it et al.} Phys. Rev. Lett. \bf{ 89}, 231301 (2002); 
O. Bertolami, Gen. Relativ. Gravit. \bf {34}, 707 (2002);
T. Jacobson {\it et al.}, Nature (London) \bf {424},
1019 (2003); J. Alfaro and G. Palma, Phys. Rev. D \bf {67}, 083003 (2003);
R. C. Myers and M. Pospelov, gr-qc/0402028; 

\bibitem{Coleman:1998ti} 
S.~R.~Coleman and S.~L.~Glashow,
Phys.\ Rev.\ D \bf {59}, 116008 (1999),
hep-ph/9812418.

\bibitem{GrootNibbelink:2004za}
S. Groot Nibbelink and M. Pospelov,
hep-ph/0404271.


\bibitem{Carmona:2000gd}
J.~M.~Carmona and J.~L.~Cortes,
Phys.\ Rev.\ D {\bf 65}, 025006 (2002), hep-th/0012028.

\bibitem{Wess} J. Wess and B. Zumino, Phys. Lett. \bf {B49}, 52 (1974);
Nucl. Phys. \bf {B70}, 39 (1974).


\bibitem{Kleppe}  G. Kleppe and R.P. Woodard 
Phys. Lett. \bf {B253}, 331 (1991).

\bibitem{canonical} See for example, D.G. Barci, L.E. Oxman and M. Rocca
Int. J. Mod. Phys. \bf {A11}, 2111 (1996); hep-th/9503101

\bibitem{Ralston:2003pf}
P.~Jain and J.~P.~Ralston,
Mod.\ Phys.\ Lett.\ A \bf {14}, 417 (1999), astro-ph/9803164;  
J.~P.~Ralston and P.~Jain,
Int.\ J.\ Mod.\ Phys.\ D \bf{ 13}, 1857 (2004), astro-ph/0311430.

\bibitem{Anderson} See e.g. P. W. Anderson, {\it Basic Notions of 
Condensed Matter Physics}, (Addison-Wesley, 1984).

\end{thebibliography}
\end{document}